\documentclass[12pt]{article}
\usepackage{lscape}
\usepackage{citesort}
\usepackage{amssymb}
\usepackage{appendix}
\usepackage{multirow}

\begin{document}

\def\Tr{\mbox{Tr}}
\def\figt#1#2#3{
        \begin{figure}
        $\left. \right.$
        \vspace*{-2cm}
        \begin{center}
        \includegraphics[width=10cm]{#1}
        \end{center}
        \vspace*{-0.2cm}
        \caption{#3}
        \label{#2}
        \end{figure}
	}
	
\def\figb#1#2#3{
        \begin{figure}
        $\left. \right.$
        \vspace*{-1cm}
        \begin{center}
        \includegraphics[width=10cm]{#1}
        \end{center}
        \vspace*{-0.2cm}
        \caption{#3}
        \label{#2}
        \end{figure}
                }

\def\ds{\displaystyle}
\def\beq{\begin{equation}}
\def\eeq{\end{equation}}
\def\bea{\begin{eqnarray}}
\def\eea{\end{eqnarray}}
\def\beeq{\begin{eqnarray}}
\def\eeeq{\end{eqnarray}}
\def\ve{\vert}
\def\vel{\left|}
\def\ver{\right|}
\def\nnb{\nonumber}
\def\ga{\left(}
\def\dr{\right)}
\def\aga{\left\{}
\def\adr{\right\}}
\def\lla{\left<}
\def\rra{\right>}
\def\rar{\rightarrow}
\def\lrar{\leftrightarrow}  
\def\nnb{\nonumber}
\def\la{\langle}
\def\ra{\rangle}
\def\ba{\begin{array}}
\def\ea{\end{array}}
\def\tr{\mbox{Tr}}
\def\ssp{{\Sigma^{*+}}}
\def\sso{{\Sigma^{*0}}}
\def\ssm{{\Sigma^{*-}}}
\def\xis0{{\Xi^{*0}}}
\def\xism{{\Xi^{*-}}}
\def\qs{\la \bar s s \ra}
\def\qu{\la \bar u u \ra}
\def\qd{\la \bar d d \ra}
\def\qq{\la \bar q q \ra}
\def\GG{\langle g_s^2 G^2 \rangle}
\def\g5{\gamma_5 \not\!q}
\def\x{\gamma_5 \not\!x}
\def\g5{\gamma_5}
\def\sb{S_Q^{cf}}
\def\sd{S_d^{be}}
\def\su{S_u^{ad}}
\def\sbp{{S}_Q^{'cf}}
\def\sdp{{S}_d^{'be}}
\def\sup{{S}_u^{'ad}}
\def\ssp{{S}_s^{'??}}

\def\sig{\sigma_{\mu \nu} \gamma_5 p^\mu q^\nu}
\def\fo{f_0(\frac{s_0}{M^2})}
\def\ffi{f_1(\frac{s_0}{M^2})}
\def\fii{f_2(\frac{s_0}{M^2})}
\def\O{{\cal O}}
\def\sl{{\Sigma^0 \Lambda}}
\def\es{\!\!\! &=& \!\!\!}
\def\ap{\!\!\! &\approx& \!\!\!}
\def\ar{&+& \!\!\!}
\def\arrr{\!\!\!\! &+& \!\!\!}
\def\ek{&-& \!\!\!}
\def\kek{\!\!\!\!&-& \!\!\!}
\def\cp{&\times& \!\!\!}
\def\se{\!\!\! &\simeq& \!\!\!}
\def\eqv{&\equiv& \!\!\!}
\def\kpm{&\pm& \!\!\!}
\def\kmp{&\mp& \!\!\!}
\def\mcdot{\!\cdot\!}
\def\erar{&\rightarrow&}


\def\simlt{\stackrel{<}{{}_\sim}}
\def\simgt{\stackrel{>}{{}_\sim}}


\title{
         {\Large
                 {\bf
$\gamma^\ast N \to N(1535)$ transition form factors
in QCD
                 }
         }
      }

\author{\vspace{1cm}\\
{\small T. M. Aliev \thanks {
taliev@metu.edu.tr}~\footnote{Permanent address: Institute of
Physics, Baku, Azerbaijan.}\,\,,
M. Savc{\i} \thanks
{savci@metu.edu.tr}} \\
{\small Physics Department, Middle East Technical University,
06531 Ankara, Turkey }}

\date{}

\begin{titlepage}
\maketitle
\thispagestyle{empty}

\begin{abstract}

The form factors, as well as the helicity amplitudes for the $\gamma^\ast N
\to N(1535)$ transition are calculated in framework of the QCD light cone
sum rules. The contamination coming from diagonal transition is eliminated
by considering combinations of the sum rules corresponding to different
Lorentz structures. Similar procedure is applied
for determination of the residue and mass of the negative parity baryon.
Comparison of our predictions on the helicity amplitudes
with the existing theoretical and experimental results is also presented.

\end{abstract}

~~~PACS numbers: 11.55.Hx, 13.40.Gp, 14.20.Gk

\end{titlepage}

\section{Introduction}

The quark model and quantum chromodynamics predict many exotic bound states
of baryons. Few of these states have already been discovered experimentally.
Study of the properties of these states allows to understand their complex
structure. Quite fruitful experiments have been conducted in this direction
at Jefferson Laboratory.
The radiative transition of the excited to ground state baryons is described
by the electromagnetic form factors. The study of these form factors can
provide us information about the quark structures of hadrons.

In the present work we calculate the $\gamma^\ast N \to N(1535)$
electromagnetic transition form factors in framework of the light cone QCD
sum rules method (about this method, see for example \cite{Rgsn01}).
This transition is comprehensively studied
experimentally and there already accumulated rich data (for a review see
\cite{Rgsn02} and the references therein). This transition has been studied
within the covariant quark model and in lattice gauge theory in
\cite{Rgsn03} and \cite{Rgsn04}, respectively. The analysis of this
transition is interesting due to the fact that, the initial and final
baryons have different masses, as well as different parities.

The plan of the work is as follows. Section 2 is devoted to the derivation
of the sum rules for the form factors of the $\gamma^\ast N \to N(1535)$
transition within light cone QCD sum rules. In this section we also present
expressions of the helicity amplitudes. In section 3, the numerical analysis
of the sum rules for the form factors is given.

\section{Light cone QCD sum rules for the form factors of the
$\gamma^\ast N \to N(1535)$ transition}

The experimental study of the electromagnetic form factors of the excited state
baryon to the ground state baryon transition is quite useful for extracting
information about the mechanism of the strong interactions at low energies.
Theoretically these form factors appear when the electromagnetic current is
sandwiched between the initial and
final states, i.e., $\lla N^\ast(p^\prime) \vel J_\mu^{el} (q) \ver N(p)
\rra$~, where $J_\mu^{el}$ is the electromagnetic current with
four-momentum $q_\mu = (p-p^\prime)_\mu$ and, customarily, $N^\ast$ denotes
$N(1535)$, which is a negative parity baryon.
This matrix element is described with the help of two form factors as
follows:
\bea
\label{egsn01}
\lla N^\ast(p^\prime) \vel J_\mu^{el} (q) \ver N(p) \rra \es
\bar{u}_{N^\ast} (p^\prime) \Bigg[ \Bigg( \gamma_\mu - {\rlap/{q} q_\mu
\over q^2} \Bigg) F_1^\ast (Q^2) \nnb \\
\ek {i \over m_N + m_{N^\ast} }
\sigma_{\mu\nu} q^\nu F_2^\ast (Q^2) \Bigg] \gamma_5 u_N(p)~,
\eea
where $Q^2=-q^2$, $\sigma_{\mu\nu} = {i\over 2} [\gamma_\mu,\gamma_\nu]$, and
$F_1^\ast (Q^2)$ and $ F_2^\ast (Q^2)$ are the transition form factors.
This definition, obviously, satisfies the conservation of the electromagnetic
current.
In order to determine the transition form factors $F_1^\ast (Q^2)$ and $
F_2^\ast (Q^2)$ from light cone sum rules (LCSR), we start by considering
the correlation function,
\bea
\label{egsn02}
\Pi_\mu (p,q) \es i \int d^4x e^{i q x}\lla 0 \vel {\cal T} \{\eta(0) J_\mu^{el}(x)
\} \ver N(p) \rra~,
\eea
where $\eta$ is the interpolating current with the nucleon quantum numbers,
i.e.,
\bea
\label{egsn03}
\eta = 2 \varepsilon^{abc} \Big\{ (u^{aT} C d^b) \gamma_5 u^c + \beta (u^{aT} C
\gamma_5 d^b) u^c \Big\}~,
\eea
and $J_\mu^{el} = e_u \bar{u} \gamma_\mu u + e_d \bar{d} \gamma_\mu d$, $C$ is
the charge conjugation operator, and $\beta$ is an arbitrary parameter. The
choice $\beta=-1$ corresponds to the well known Ioffe current. Using
the quark-hadron duality through dispersion relations one can write
expressions for the form factors in terms of the distribution amplitudes of
the nucleon. In realizing this program, we first calculate the correlator
(\ref{egsn01}) in terms of hadrons. Applying the standard procedure of the
QCD sum rules approach, we insert the ``full" set of hadronic states between
the currents $\eta$ and $J_\mu^{el}$ in Eq. (\ref{egsn02}). As the result,
the contributions of the lowest nucleon state and its negative parity
partner $N^\ast$ enter to the correlator and we get,
\bea
\label{egsn04}
\Pi_\mu (p,q) \es { \lla 0 \vel \eta \ver N(p^\prime) \rra \lla N(p^\prime)
\vel J_\mu^{el} \ver N(p) \rra \over m_N^2 -p^{\prime 2} } +
{ \lla 0 \vel \eta \ver N^\ast(p^\prime) \rra \lla N^\ast(p^\prime) 
\vel J_\mu^{el} \ver N(p) \rra \over m_{N^\ast}^2 -p^{\prime 2} }~.
\eea
The couplings of $N^\ast$ and $N$ with the interpolating current $\eta$ are
defined in the following way:
\bea          
\label{egsn05}
\lla 0 \vel \eta \ver N(p^\prime) \rra \es \lambda_N u_N (p^{\prime})~, \nnb \\
\lla 0 \vel \eta \ver N^\ast(p^\prime) \rra \es \lambda_{N^\ast} \gamma_5 u_{N^\ast}
(p^{\prime})~.
\eea
The matrix element $\lla N^\ast(p^\prime) \vel J_\mu^{el} \ver N(p) \rra$ in
terms of the form factors $F_1^\ast (Q^2)$ and $ F_2^\ast (Q^2)$ is already
given in Eq. (\ref{egsn01}). The matrix corresponding to the diagonal
$\lla N(p^\prime) \vel J_\mu^{el} \ver N(p) \rra$ transition
can be obtained from Eq. (\ref{egsn01}) by
making the replacements, $m_{N^\ast} \to m_N$, $F_1^\ast \to F_1$,
$F_2^\ast \to F_2$, and omitting $\gamma_5$ and ${\rlap/{q} q_\mu
/q^2}$. Using Eqs. (\ref{egsn01}) and (\ref{egsn05}), and performing
summation over over the spins of $N$ and $N^\ast$, from Eq.
(\ref{egsn04}) we get
\bea          
\label{egsn06}
\Pi_\mu (p,q) \es {1 \over m_N^2 -p^{\prime 2} } \lambda_N
({\not\!{p}}^\prime + m_N) \Bigg[ \gamma_\mu F_1(Q^2)  - {i \over 2 m_N}
\sigma_{\mu\nu} q^\nu F_2(Q^2) \Bigg] u_N (p) \nnb \\
\ar {1 \over m_{N^\ast}^2 -p^{\prime 2} } \lambda_{N^\ast} \gamma_5
({\not\!{p}}^\prime + m_{N^\ast}) \Bigg[ \Big(\gamma_\mu - {\rlap/{q} q_\mu
\over q^2}\Big)  F_1^\ast(Q^2)  - {i \over m_N+m_N^\ast}
\sigma_{\mu\nu} q^\nu F_2^\ast (Q^2) \Bigg] \gamma_5 u_N (p)~,
\eea
where $q=p-p^\prime$.
Using the Dirac equation $(\not\!{p} - m_N) u_N(p)=0$, and ordering the
Dirac matrices as $\rlap/{q} \gamma_\mu$ we get from Eq.
(\ref{egsn06}) the following expressions for the invariant amplitudes for
the structures $\gamma_\mu$, $q_\mu$ and $q_\mu \rlap/{q}$, respectively,
\bea          
\label{egsn07}
\Pi_1((p-q)^2, q^2) \es - {\lambda_{N^\ast} (m_{N^\ast}+m_N) \over
m_{N^\ast}^2 - (p-q)^2 } F_1^\ast (q^2) -
{\lambda_{N^\ast} (m_{N^\ast}-m_N) \over
m_{N^\ast}^2 - (p-q)^2 } F_2^\ast (q^2) \nnb \\
\ar \int_{s_0}^\infty ds {\rho_1(s,q^2) \over s-(p-q)^2 }~, \\ \nnb \\
\label{egsn08}
\Pi_2((p-q)^2, q^2) \es {\lambda_{N^\ast} (m_{N^\ast}^2 - m_N^2) \over q^2[m_{N^\ast}^2 -
(p-q)^2] } F_1^\ast (q^2) +
{\lambda_{N^\ast} (m_{N^\ast}-m_N) \over
(m_{N^\ast} + m_N) [m_{N^\ast}^2 -(p-q)^2] } F_2^\ast (q^2) \nnb \\
\ar {\lambda_N \over m_N^2 - (p-q)^2} F_2(q^2) +
\int_{s_0}^\infty ds {\rho_2(s,q^2) \over s-(p-q)^2 }~, \\ \nnb \\
\label{egsn09}
\Pi_3((p-q)^2, q^2) \es {\lambda_{N^\ast} (m_{N^\ast} + m_N) \over q^2[m_{N^\ast}^2 -
p^{\prime 2}] } F_1^\ast (q^2) +
{\lambda_{N^\ast} \over (m_{N^\ast}+m_N)[m_{N^\ast}^2 - (p-q)^2] }
F_2^\ast (q^2) \nnb \\
\ek {\lambda_N \over 2 m_N [m_N^2 - (p-q)^2]} F_2(q^2) +
\int_{s_0}^\infty ds {\rho_3(s,q^2) \over s-(p-q)^2 }~,
\eea
where the contributions of all excited states and continuum with the quantum
numbers of $N$ and $N^\ast$ are taken into account using the quark-hadron
duality ansatz.

Having obtained the expression of the correlator from hadronic side, we
proceed now to calculate it in terms of quark and gluon degrees of freedom,
in which $(p-q)^2$ and $q^2=-Q^2$ are large and negative, i.e., they are
Eucledian. This is necessary for justifying the operator product expansion
(OPE) of the two currents around the light cone. The expression for the
correlator is derived in terms of nucleon distribution amplitudes (DAs)
with increasing twist, convoluted with hard-scattering amplitudes. Equating
then the coefficients of the structures $\gamma_\mu$, $q_\mu$ and $q_\mu
\rlap/{q}$ of the correlators from hadronic and QCD sides we obtain the sum
rules for the combinations of the transition form factors $F_1^\ast(Q^2)$,
$F_2^\ast(Q^2)$ and $F_2(Q^2)$. The nucleon DAs from twist-3 to twist-6
are obtained in \cite{Rgsn05}, and we use these DAs in further
numerical analysis.

Using the expressions of the DAs from correlator (\ref{egsn01}) for the
invariant amplitudes of the structures $\gamma_\mu$, $q_\mu$ and $q_\mu
\rlap/{q}$, we get
\bea          
\label{egsn10}
\Pi_i^{th} ((p-q)^2,q^2) = \sum_{n=1}^{3} \int_0^1 {\rho_{i,n}
(x,(p-q)^2,q^2) \over [(q-px)^2]^n }~.
\eea
After tedious calculations we obtain the expressions for the functions
$\rho_{i,n}$, where  $i$ and $n$ run through 1,2,3,
and their explicit forms are all given in the Appendix.

Equating Eqs. (\ref{egsn07}), (\ref{egsn08}) and (\ref{egsn09}) with
$\Pi_1^{th}$, $\Pi_2^{th}$ and $\Pi_3^{th}$, from Eq. (\ref{egsn10}) we get
the sum rules for the combination of the form factors. Eliminating the
$F_2(Q^2)$ from this set of equations, and performing the Borel
transformation on the variable $(p-q)^2 \to M^2$, we get the following sum
rules for the combinations of the form factors $F_1^\ast(Q^2)$ and 
$F_2^\ast(Q^2)$ as follows:
\bea
\label{egsn11}
\lambda_{N^\ast} \Big[ (m_{N^\ast} + m_N) F_1^\ast(Q^2) + (m_{N^\ast} - m_N)
F_2^\ast(Q^2) \Big] e^{-m_{N^\ast}^2/M^2} \es - {\cal J}_1(Q^2,M^2,s_0)~, \nnb \\
\lambda_{N^\ast} \Bigg[ - {(m_N + m_{N^\ast})^2 \over Q^2} F_1^\ast(Q^2) +
F_2^\ast(Q^2) \Bigg] e^{-m_{N^\ast}^2/M^2} \es {\cal J}_2 (Q^2,M^2,s_0) \nnb \\
\ar 2 m_N {\cal J}_3 (Q^2,M^2,s_0)~,
\eea
where $M^2$ is the Borel mass parameter, $s_0$ is the continuum threshold, and
\bea
\label{nolabel01}
s \es {1-x \over x} Q^2 + (1-x) m_N^2~, \nnb \\
x_0 \es {1\over 2 m_N^2} \Big[ \sqrt{(Q^2 + s_0 - m_N^2)^2 +4 m_N^2 Q^2} -
(Q^2+s_0-m_N^2) \Big]~. \nnb
\eea
The functions ${\cal J}_{i}(Q^2,M^2,s_0)$ appearing on the right-hand side of Eq.
(\ref{egsn11}) have the following form: 
\bea
\label{egsn12} {\cal J}_{i}(Q^2,M^2,s_0) \es
\int_{x_0}^1 dx \Bigg[
- {\rho_{i,1}(x) \over x} + {\rho_{i,2}(x) \over x^2 M^2} \nnb \\
\ek {\rho_{i,3}(x) \over
2 x^3 M^4}\Bigg] e^{-\left({\bar{x} Q^2 \over x M^2} + {\bar{x} m_N^2 \over
M^2}\right)} +
\Bigg[ {\rho_{i,2}(x_0) \over Q^2 + x_0^2 m_N^2} - {1 \over 2 x_0}
{\rho_{i,3}(x_0) \over (Q^2 + x_0^2 m_N^2) M^2} \nnb \\
\ar {1\over 2} {x_0^2 \over Q^2 + x_0^2 m_N^2} {d\over dx_0} {\rho_{i,3}(x_0)
\over x_0(Q^2 + x_0^2 m_N^2) M^2} \Bigg] e^{-s_0/M^2}~.
\eea
Explicit expressions of the functions $\rho_{i,n}$ are presented in the
Appendix.

From experimental point of view the helicity amplitudes seem to be more
suitable quantities for the analysis of
$\gamma^\ast N \to N(1535)$ transition. the helicity amplitudes are
determined in terms of the transition form factors as (see \cite{Rgsn07})
\bea
\label{egsn13}
A_{1/2} \es - 2 e \sqrt{ {(m_{N^\ast} + m_N)^2 +Q^2 \over 8 m_N (m_{N^\ast}^2
- m_N^2) } } \Bigg[ F_1^\ast (Q^2) + {m_{N^\ast} - m_N \over m_{N^\ast} + m_N}
F_2^\ast (Q^2) \Bigg]~, \\ \nnb \\
\label{egsn14}
S_{1/2} \es \sqrt{2} e \sqrt{ {(m_{N^\ast} + m_N)^2 +Q^2 \over 8 m_N
(m_{N^\ast}^2 - m_N^2) } } ( m_{N^\ast} + m_N )
{\sqrt{\lambda(Q^2,m_{N^\ast}^2,m_N^2)} \over 2 m_{N^\ast} Q^2}
\Bigg[ {m_{N^\ast} - m_N \over m_{N^\ast} + m_N}  F_1^\ast (Q^2) \nnb \\
\ek  {Q^2 \over (m_{N^\ast} + m_N)^2 } F_2^\ast (Q^2) \Bigg]~,
\eea
where,
\bea
\label{nolabel04}
\lambda(Q^2,m_{N^\ast}^2,m_N^2) = (Q^2+m_{N^\ast}^2+m_N^2)^2 -
4 m_{N^\ast}^2 m_N^2~. \nnb
\eea
So, from Eqs. (\ref{egsn11}) and (\ref{egsn12}) we can find $Q^2$ dependence
of the transition form factors $F_1^\ast(Q^2)$ and $F_2^\ast(Q^2)$. Then
using Eqs. (\ref{egsn13}) and (\ref{egsn14}) we can make predictions about
the helicity amplitudes and compare these results with the present
experimental results.

It follows from Eq. (\ref{egsn11}) that, in order to calculate the transition
form factors $F_1^\ast(Q^2)$ and
$F_2^\ast(Q^2)$ the residue of the $N^\ast$ baryon is needed. For this
purpose we consider the two-point correlation function
\bea
\label{egsn15} 
\Pi(p^2) = i \int d^4x e^{ipx} \lla 0 \vel {\cal T} \{ \eta(x)
\bar{\eta}(0) \} \ver 0 \rra~.
\eea
This correlation function contain two invariant amplitudes, namely,
coefficients of the structures $\not\!{p}$ and the unit operator $I$, which
can be written as
\bea
\label{nolabel05}
\Pi(p^2) = \Pi_1 (p^2) \!\not\!{p} + \Pi_2 (p^2) I~. \nnb
\eea
The theoretical part of this correlation function has already been
calculated in \cite{Rgsn06} using the general form of the interpolating
current of the $N$ baryon which is given in Eq. (\ref{egsn03}).
Saturating  the correlation function with $N$ and
$N^\ast$ states, we get
\bea
\label{egsn16}
\Pi(p^2) = {\vel \lambda_{N^\ast} \ver^2 \over  m_{N^\ast}^2 - p^2 }
(\not\!{p} - m_{N^\ast} ) + {\vel \lambda_N \ver^2 \over
m_N^2 - p^2 } (\not\!{p} + m_N ) + \cdots~,
\eea
where $\cdots$ denotes the contributions of the higher states and
continuum. Eliminating the contributions of the $N$ states and performing
Borel transformation for the mass and residue of the negative parity baryon
we get the following sum rules:
\bea
\label{egsn17}
m_{N^\ast}^2 \es { \ds \int_0^{s_0} ds \, e^{-s/M^2} s \Big[ m_N \mbox{Im}
\Pi_1(s) - \mbox{Im} \Pi_2 (s) \Big] \over
\ds \int_0^{s_0} ds \, e^{-s/M^2} \Big[ m_N \mbox{Im}
\Pi_1(s) - \mbox{Im} \Pi_2 (s) \Big] }~, \\ \nnb \\
\label{egsn18}
\vel \lambda_{N^\ast} \ver^2 \es {e^{m_{N^\ast}^2/M^2} \over m_{N^\ast} + m_N}
{1\over \pi} \int_0^{s_0} ds \,e^{-s/M^2} \Big[ m_N \mbox{Im}
\Pi_1(s) - \mbox{Im} \Pi_2 (s) \Big]~.
\eea
As has already been noted the functions $\mbox{Im}
\Pi_1(s)$ and $\mbox{Im} \Pi_2 (s)$ are calculated in \cite{Rgsn06}, and
therefore we do not present their expressions here. Substituting Eq.
(\ref{egsn17}) into Eq. (\ref{egsn18}) we can determine the residue of
$N^\ast$.

\section{Numerical analysis}
In this section we present numerical analysis of the obtained sum rules for
the helicity amplitudes. 
The main input parameters of the light cone sum rules analysis are the DAs
of nucleon, which are all presented in the Appendix. The normalization
factors appearing in them are determined from the analysis of two-point sum
rules \cite{Rgsn05} which are predicted to have the values
\bea
\label{nolabel07}
f_N       \es   (5.0\pm 0.5)\times 10^{-3}~\mbox{GeV}^2~, \nnb \\
\lambda_1 \es - (2.7 \pm 0.9)\times 10^{-2}~\mbox{GeV}^2~, \nnb \\
\lambda_2 \es - (5.4 \pm 1.9)\times 10^{-2}~\mbox{GeV}^2~. \nnb
\eea
For the remaining five parameters which determine the shapes of the nucleon
DAs, we use the values $A_1^u=0.13$, $V_1^d=0.30$, $f_1^d=0.33$, $f_1^u=0.09$
and $f_2^d=0.25$ that are given in \cite{Rgsn08}, and the quark condensate is
predicted to have the value $\qq (1~\mbox{GeV}) =
-(246_{-19}^{+28}~\mbox{MeV})^3$
\cite{Rgsn09}.

In order to determine the transition form factors we need to know the mass
and residue of $N^\ast$, which can be obtained from the analysis of the
two-point sum rules (see Eq. (\ref{egsn17}) into Eq. (\ref{egsn18})).
These sum rules contain two auxiliary parameters, one being the Borel mass
parameter and the other is the continuum threshold $s_0$. 
The working domain of the Borel mass parameter
$M^2$ is determined by means of the standard criteria, namely, the
nonperturbative and continuum contributions in the sum rules should
sufficiently be suppressed. The value of the threshold $s_0$ is chosen in
such a way that, the sum rules prediction reproduces the experimentally
measured mass with high enough accuracy, say within 10\%.

The value of the residue of $N^\ast$ can be calculated in a way as exploited
below. At the first stage we analyze the mass sum rule at several different values of the
arbitrary parameter $\beta$, and observe that the aforementioned criteria is
satisfied well in the range $1.0\le M^2 \le 3.0~\mbox{GeV}^2$ of the Borel mass
parameter. We also see that for the choice of the continuum threshold 
in the range $s_0=(4.0 \pm 0.5)~\mbox{GeV}^2$ the sum rules reproduce the mass of
the $N^\ast$ baryon with the limits of required accuracy. The second step
in the sum rules analysis is determination of the working region of the
parameter $\beta$, where the sum rules prediction of the mass is
independent of it and reproduces the measured mass of $N^\ast$. We
further demand that $\lambda_{N^\ast}^2$ is also independent of $\beta$ and
positive in the domain we are looking for. Our calculations show that both
conditions are fulfilled in the domain $-0.4 \le \beta \le 0.8$, and  we
shall use it in further analysis of the sum rules for the transition form
factors.

In Figs. (1) and (2), the dependencies of the helicity amplitudes $A_{1/2}$
and $S_{1/2}$ on $Q^2$ at $M^2=2.0~\mbox{GeV}^2$, at several fixed values of the
auxiliary parameter $\beta$ picked from its working region are presented,
respectively. Note that the data in Figs. (1) and (2) are obtained only for
the central values of the input parameters that enter to the sum rules for
the form factors $F_1^\ast (Q^2)$ and $F_2^\ast (Q^2)$. The numerical
analysis shows that the results for the above-mentioned helicity amplitudes
are insensitive to the variation of in $M^2$ in the region $1.4 \le
M^2 \le 2.0~\mbox{GeV}^2$.  For completeness,
in these figures the experimental data from CLAS
\cite{Rgsn10,Rgsn11} and MAID \cite{Rgsn12}  are also presented.

We see from Fig. (1) that the value of $A_{1/2}$ decreases with the
increasing values of $\beta$. This behavior is to the contrary in the case
of $S_{1/2}$, i.e., $S_{1/2}$ increases along with the decreasing values of
$\beta$. In other words, $A_{1/2}$ and $S_{1/2}$ both seem to be quite
sensitive to the values of the auxiliary parameter $\beta$. It follows from 
these figures that when the central values of the input parameters are taken
into account, the values of $A_{1/2}(S_{1/2})$ are larger (smaller) compared
to the experimental data, as well as the lattice result \cite{Rgsn08},
in the appropriate
working region of the auxiliary parameter $\beta$. It is also observed that
with the increasing(decreasing) value of $\beta$ the difference between our
prediction on $A_{1/2}(S_{1/2})$ 
and the experimental data and lattice result gets smaller and smaller, and
they practically coincide with each other when the the parameter
$\beta$ is around $0.6(-0.3)$.

If the uncertainties in the input parameters are taken into account we
observe that our predictions and the experimental data on the helicity
amplitudes are very close to each other when $\beta$ ranges in the domain
$-0.1 \le \beta \le 0.4$, and they are in good agreement  with the
lattice results \cite{Rgsn08}. Small differences in our results and
\cite{Rgsn08} can be attributed to the errors in the value of the residue of
$N^\ast$, as well as to the values of the DAs, since different sets of DAs
have been used in \cite{Rgsn08}, while DAs of $N$ have been used in
our work.So, we observe the experimental data, lattice results and our
predictions coincide only for a more restricted domain of the arbitrary
parameter $\beta$ in the range $-0.1 \le \beta \le 0.4$. Therefore, we can
conclude that the working region $-0.6 \le \beta \le 0.8$ of the arbitrary
parameter, which follows from the analysis of the two-point sum rules,
is rather restricted into the domain $-0.1 \le \beta \le 0.4$.

Finally, when we compare our predictions with those of the
covariant quark model \cite{Rgsn03}, we observe that our results for the
helicity amplitude $A_{1/2}$ are considerably larger in magnitude.
Furthermore, in the case of $S_{1/2}$ our results are close to those given
in \cite{Rgsn03} within the error limits only for $\beta=0.1$, while they
differ considerably for all other values of $\beta$. More refined
calculations of the DAs and of the radiative corrections to them could of
course allow further improvements in our results. For a reliable
comparison of the theoretical results with the experiments more effort
from both sides are needed.

In conclusion, the helicity amplitudes of the  $\gamma N \to N^\ast(1535)$
transition are calculated within the light cone QCD sum rules method, where
the DAs of the nucleon are used. The unwanted contribution
coming from the  diagonal transition is eliminated by considering
combination of the sum rules. We also presented a
comparison of our predictions on the helicity amplitudes with the theoretical
and experimental results existing in the literature.
From these comparisons we see that our predictions on the helicity
amplitudes $A_{1/2}$ and $S_{1/2}$ agree well with the experimental data and
lattice results within the error limits in the input parameters, and only
for the small values of $\beta$ which is restricted into the range $-0.1 \le
\beta \le 0.4$. This restriction is more stringent compared to the one
imposed by the mass sum rule analysis, i.e., $-0.4 \le \beta \le 0.8$. The
result obtained for $\beta$ can also be used in the analysis of the
transition form factors for the other members the negative parity octet
baryons.

\newpage

\section*{Appendix}  
\setcounter{equation}{0}
\setcounter{section}{0}

In this Appendix we present the explicit expressions of the functions
$\rho_{i,n}$ entering into the sum rules.
\section{Structure $\gamma_\mu$}
%
%
%
%
\bea
%
%
\rho_{1,1} (x)\es
%
%
- {1\over 2 x} e_{q_1} m_N \Bigg\{ m_{q_1} \Big[ 2 (1+\beta) (\check{C}_2         
+ \check{D}_2) + (1-\beta) (\check{B}_2 + 5 \check{B}_4) \Big] \nnb \\
\ek 2 x (1+\beta) \int_0^{\bar{x}} dx_3 \Big[m_N x            
( P_1 + S_1 + 3 T_1 - 6 T_3) \nnb \\
\ek m_{q_1} ( A_1 + 2 A_3 - V_1 + 2 V_3) \Big]
(x,1-x-x_3,x_3) \nnb \\
\ar 2 x (1-\beta) \int_0^{\bar{x}} dx_3 \Big[m_N x (A_1 + 2 A_3 - V_1 +2 V_3) \nnb \\
\ek  m_{q_1} ( P_1 + S_1 + 3 T_1 - 6 T_3) \Big] (x,1-x-x_3,x_3) \Bigg\} (x)\nnb \\
%
%
\ar {1\over 2 x} e_{q_2} \Bigg\{ m_N^2 (1+\beta) \; \widetilde{\!\widetilde{B}}_6 \nnb \\
\ar  2 m_N (1+\beta) \Big[ m_N x (\widetilde{B}_4
+ \widetilde{B}_5 + 2 \widetilde{B}_7 + \widetilde{E}_1 - \widetilde{H}_1 ) 
+ m_{q_2} (\widetilde{C}_2 +\widetilde{D}_2) \Big] \nnb \\
\ar m_N (1-\beta) \Big[ 2 m_N x (\widetilde{C}_2 - \widetilde{C}_5 -
\widetilde{D}_2 - \widetilde{D}_5) - m_{q_2} (\widetilde{B}_2               
+\widetilde{B}_4) \Big] \nnb \\
\ar 2 m_N x (1+\beta) \int_0^{\bar{x}} dx_1 \Big[m_N x 
( P_1 + S_1 + T_1 - 2 T_3)  \nnb \\
\ar m_{q_2} ( A_1 + A_3 - V_1 + V_3) \Big]
(x_1,x,1-x-x_1) \nnb \\
\ar 2 m_N (1-\beta) \int_0^{\bar{x}} dx_1 \Big[m_N (A_1^M + T_1^M) + Q^2 (A_1 + V_1) \nnb \\
\ar x m_{q_2} (P_1 - S_1 + T_1) \Big] (x_1,x,1-x-x_1) \Bigg\} (x) \nnb \\
%
%
\ek {1\over 2 x} e_{q_3} \Bigg\{ 2 m_N^2 (1+\beta) \; \widehat{\!\widehat{B}}_6 \nnb \\
\ek m_N (1+\beta) \Big[ 2 m_N x
(\widehat{B}_2 +
\widehat{B}_4 + 2 \widehat{B}_5 + 4 \widehat{B}_7)
- m_{q_3} (\widehat{B}_2 +\widehat{B}_4) \Big] \nnb \\
\ek m_N (1-\beta) \Big[ m_N x (\widehat{C}_4 - \widehat{C}_5 -
\widehat{D}_4 + \widehat{D}_5) + 2 m_{q_3} (\widehat{B}_2
+\widehat{B}_4) \Big] \\
\ek 2 m_N x (1+\beta) \int_0^{\bar{x}} dx_1 \Big[m_N x 
( P_1 + S_1 - T_1 + 2 T_3) \nnb \\
\ek m_{q_3} ( P_1 - S_1 - T_1) \Big]
(x_1,1-x-x_1,x) \nnb \\
\ek 2 (1-\beta) \int_0^{\bar{x}} dx_1 \Big[m_N^2 (A_1^M - V_1^M) + Q^2 (A_1 - V_1)
\nnb \\
\ek x m_{q_3} m_N (A_1 + A_3 + V_1- V_3) \Big] (x_1,1-x-x_1,x)\Bigg\} (x)~.
\nnb \\ \nnb \\ \nnb \\
%
%
%
\rho_{1,2} (x)\es
%
%
- {1\over 2x} e_{q_1} m_N \Bigg\{2 m_N^3 x^2 (1+\beta) (\, \check{\!\check{B}}_6
- 3 \, \check{\!\check{B}}_8) - m_N^2 x (1-\beta)\Big[ m_{q_1} \,
\check{\!\check{B}}_6 + 2 m_N x (\, \check{\!\check{C}}_6 
+ \, \check{\!\check{D}}_6 ) \Big] \nnb \\
\ar (1+\beta) \Big[ m_N x (m_N^2 x^2 +Q^2)      
(\check{B}_2 + 5 \check{B}_4) \nnb \\
\ek x^2 m_{q_1} m_N^2 ( \check{C}_4 - 3
\check{C}_5 - \check{D}_4 + 3 \check{D}_5 ) + 2 m_{q_1} Q^2( \check{C}_2 +
\check{D}_2) \Big] \nnb \\
\ar (1-\beta) \Big[ 2 m_N x (m_N^2 x^2 +Q^2) ( \check{C}_2
+ \check{D}_2) + m_{q_1} Q^2 (\check{B}_2 + 5 \check{B}_4)\nnb \\
\ek x^2 m_{q_1} m_N^2 (\check{B}_2 - \check{B}_4
+ 6 \check{B}_5 + 12 \check{B}_7 - 2 \check{E}_1 + 2 \check{H}_1)
\Big]\nnb \\
\ek 2 m_N^3 x^2  \int_0^{\bar{x}} dx_3 \Big[  3 (1+\beta) T_1^M -
(1-\beta) ( A_1^M - V_1^M ) \Big] (x,1-x-x_3,x_3)\Bigg\} (x) \nnb \\
%
%
\ar {1\over 2 x} e_{q_2} m_N \Bigg\{ m_N (1+\beta) \Big[ Q^2 \;
\widetilde{\!\widetilde{B}}_6 - m_N^2 x^2 (\; \widetilde{\!\widetilde{B}}_6    
- 2\; \widetilde{\!\widetilde{B}}_8) - 2 m_{q_2} m_N x 
(\; \widetilde{\!\widetilde{C}}_6 + \; \widetilde{\!\widetilde{D}}_6) \Big] \nnb \\
\ar m_N^2 x (1-\beta) \Big[ 2 m_N x (\; \widetilde{\!\widetilde{C}}_6 - 
\; \widetilde{\!\widetilde{D}}_6) + m_{q_2} m_N (\;\widetilde{\!\widetilde{B}}_6 - 
4 \; \widetilde{\!\widetilde{B}}_8) \Big] \nnb \\
\ek (1+\beta) \Big[ m_N x (m_N^2 x^2 +Q^2)  
(\widetilde{B}_2 + \widetilde{B}_4) \nnb \\
\ar x^2 m_{q_2} m_N^2 ( \widetilde{C}_4 - \widetilde{C}_5 -
\widetilde{D}_4 + \widetilde{D}_5 ) + 2 m_{q_2} Q^2 ( \widetilde{C}_2 +
\widetilde{D}_2) \Big] \nnb \\
\ar (1-\beta) m_{q_2} \Big[ Q^2 (\widetilde{B}_2 + \widetilde{B}_4) -
m_N^2 x^2 (\widetilde{B}_2 - \widetilde{B}_4 + 2 \widetilde{B}_5 + 2
\widetilde{E}_1 + 2 \widetilde{H}_1) \Big] \nnb \\
\ar 2 m_N \int_0^{\bar{x}} dx_1 \Big[ m_N^2 x^2 (1+\beta) T_1^M +
Q^2 (1-\beta) ( A_1^M + V_1^M ) \Big] (x_1,x,1-x_1-x) \Bigg\} (x) \nnb \\
%
%
\ek {1\over 2 x} e_{q_3} m_N \Bigg\{
m_N (1+\beta) \Big[ 2 (Q^2 \; \widehat{\!\widehat{B}}_6 +
m_N^2 x^2 \; \widehat{\!\widehat{B}}_8 ) + m_{q_3} m_N x (\;
\widehat{\!\widehat{B}}_6 - 4 \; \widehat{\!\widehat{B}}_8) \Big] \nnb \\
\ar 2 m_N^2 x (1-\beta) \Big[ m_N x (     
\; \widehat{\!\widehat{C}}_6 + \; \widehat{\!\widehat{D}}_6 ) +
m_{q_3} (\; \widehat{\!\widehat{C}}_6 - \; \widehat{\!\widehat{D}}_6  
)\Big] \nnb \\
\ek (1+\beta) \Big[ m_N x (m_N^2 x^2 +Q^2)
(\widehat{B}_2 + \widehat{B}_4) - m_{q_3} Q^2 (\widehat{B}_2 +
\widehat{B}_4) \nnb \\
\ek x^2 m_{q_3} m_N^2 ( \widehat{B}_2 -
\widehat{B}_4 + 2 \widehat{B}_5 - 2 \widehat{E}_1 - 2 \widehat{H}_1) \Big]\nnb \\
\ek (1-\beta) m_{q_3} \Big[ 2 Q^2 (\widehat{C}_2 - \widehat{D}_2) -
m_N^2 x^2 (\widehat{C}_4 - \widehat{C}_5 + \widehat{D}_4 - \widehat{D}_5
\Big] \nnb \\
\ar 2 m_N \int_0^{\bar{x}} dx_1 \Big[ m_N^2 x^2
(1+\beta) T_1^M -
Q^2 (1-\beta) ( A_1^M - V_1^M ) \Big] (x_1,1-x_1-x,x) \Bigg\} (x)~, \nnb \\ \nnb \\
\rho_{1,3} (x)\es
%
%
%
e_{q_1} m_{q_1} m_N^3 (1-\beta) (m_N^2 x^2 +Q^2) \, \check{\!\check{B}}_6 (x)\nnb \\
%
%
\ar e_{q_2} m_{q_2} m_N^3 (1-\beta) (m_N^2 x^2 +Q^2) \,
\widetilde{\!\widetilde{B}}_6 (x) \nnb\\
%
%
\ar e_{q_3} m_{q_3} m_N^3 (1+\beta) (m_N^2 x^2 +Q^2) \,
\widehat{\!\widehat{B}}_6 (x)~. \nnb
\eea
%
%
\section{Structure $q_\mu$}
\bea
%
%
\rho_{2,1} (x)\es
%
%
-{1\over x} e_{q_1} m_N \Bigg\{ (1+\beta) (\check{B}_2
+ 5 \check{B}_4) + 2 (1-\beta) (\check{C}_2 + \check{D}_2) \nnb \\
\ek 2 x  \int_0^{\bar{x}} dx_3 \Big[(1+\beta)
( P_1 + S_1 + 3 T_1 - 6 T_3) \nnb \\
\ek (1-\beta) (A_1 + 2 A_3 - V_1 + 2 V_3) \Big](x,1-x-x_3,x_3) \Bigg\} (x) \nnb \\
%
%
\ek {2\over x} e_{q_2} m_N \Bigg\{ (1+\beta) (\widetilde{B}_4 +                      
\widetilde{B}_4) \nnb \\
\ek x \int_0^{\bar{x}} dx_1 \Big[ 2 (1+\beta) (T_1 - 2 T_3) + 
(1-\beta) (A_3 - V_3) \Big] (x_1,x,1-x-x_1) \Bigg\} (x) \nnb \\
%
%
\ar {1\over x} e_{q_3} m_N \Bigg\{ (1+\beta)
(\widehat{B}_2 - 3 \widehat{B}_4) - 
2 (1-\beta) (\widehat{C}_2 + \widehat{D}_2 ) \nnb \\
\ar 2 x \int_0^{\bar{x}} dx_1 \Big[ (1+\beta)
( P_1 + S_1 + T_1 - 2 T_3) \nnb \\
\ek (1-\beta) (A_1 + A_3 - V_1 + V_3) \Big] (x_1,1-x-x_1,x) \Bigg\} (x)~.
\nnb \\ \nnb \\ \nnb \\
%
%
\rho_{2,2} (x)\es
%
%
- {1\over x} e_{q_1} m_N \Bigg\{ m_N^2 x \Big[ (1+\beta) (5\, \check{\!\check{B}}_6     
- 6 \, \check{\!\check{B}}_8) - 2 (1-\beta) (\, \check{\!\check{C}}_6 +
\, \check{\!\check{D}}_6 ) \Big] \nnb \\
\ar (1+\beta) \Big[ Q^2 (\check{B}_2
+ 5 \check{B}_4) - m_N^2 x^2 (\check{B}_2 - \check{B}_4
+ 6 \check{B}_5 + 12 \check{B}_7 - 2 \check{E}_1 + 2 \check{H}_1) \Big] \nnb \\
\ek (1-\beta) \Big[ m_N^2 x^2 ( \check{C}_4 - 3 \check{C}_5 -
\check{D}_4 + 3 \check{D}_5 ) - 2 Q^2 ( \check{C}_2 + \check{D}_2)
\Big] \nnb \\
\ek 2 m_N^2 x \int_0^{\bar{x}} dx_3 \Big[  3 (1+\beta) T_1^M -
(1-\beta) (A_1^M - V_1^M ) \Big] (x,1-x-x_3,x_3) \Bigg\} (x) \nnb \\
%
%
\ek {1\over x} e_{q_2} m_N \Bigg\{ 2 m_N^2 x (1+\beta) ( \;
\widetilde{\!\widetilde{B}}_6 - 2 \; \widetilde{\!\widetilde{B}}_8)
- 2 m_N (1-\beta) \Big[2 m_{q_2} \; \widetilde{\!\widetilde{B}}_6 + m_N x
(\; \widetilde{\!\widetilde{C}}_6 - \; \widetilde{\!\widetilde{D}}_6)
\Big] \nnb \\
\ar (1+\beta) \Big[ 2 Q^2 (\widetilde{B}_2 +
\widetilde{B}_4) - 4 m_N^2 x^2 (\widetilde{B}_5 + 2 \widetilde{B}_7) +
m_N x (\widetilde{C}_4 + \widetilde{C}_5 -
\widetilde{D}_4 - \widetilde{D}_5 ) \Big] \nnb \\
\ek m_N x (1-\beta) \Big[ m_N x (\widetilde{C}_4 - \widetilde{C}_5 +
\widetilde{D}_4 - \widetilde{D}_5 ) + 2 m_{q_2} (\widetilde{B}_2 -
\widetilde{B}_4 + 2 \widetilde{B}_5)\Big] \nnb \\
\ek 4 m_N^2 x (1+\beta) \int_0^{\bar{x}} dx_1 T_1^M (x_1,x,1-x_1-x) \Bigg\}
(x) \nnb \\
%
%
\ar {1\over x} e_{q_3} m_N \Bigg\{ m_N (1+\beta) \Big[4 m_{q_2} \;
\widehat{\!\widehat{B}}_6 - m_N x (3 \; \widehat{\!\widehat{B}}_6 -
2 \; \widehat{\!\widehat{B}}_8) \Big] \nnb \\
\ar (1+\beta) \Big[ Q^2 (\widehat{B}_2 -
3 \widehat{B}_4) + m_N^2 x^2 (\widehat{B}_2 -
\widehat{B}_4 + 2 \widehat{B}_5 - 2 \widehat{E}_1 + 2 \widehat{H}_1) \nnb \\
\ar 2 x m_{q_3} m_N (\widehat{B}_2 - \widehat{B}_4 + 2 \widehat{B}_5) \Big]
- (1-\beta) \Big[ 2 m_N^2 x^2 (\widehat{C}_5 - \widehat{D}_5) + 2 Q^2
(\widehat{C}_2 + \widehat{D}_2) \nnb \\
\ar x m_{q_3} m_N (\widehat{C}_4 + \widehat{C}_5 +
\widehat{D}_4 + \widehat{D}_5)
\Big]  \nnb \\
\ar 2 m_N^2 x \int_0^{\bar{x}} dx_1 \Big[
(1+\beta) T_1^M -
(1-\beta) ( A_1^M - V_1^M ) \Big] (x_1,1-x_1-x,x) \Bigg\} (x)~, \nnb \\ \nnb \\
%
%
\rho_{2,3} (x)\es
%
%
- 4 e_{q_1} m_N^3 (1+\beta) (m_N^2 x^2 +Q^2) \, \check{\!\check{B}}_6 (x)\nnb \\
%
%
\ek {4\over x} e_{q_2} m_{q_2} m_N^2 (1-\beta)
(Q^2\; \widetilde{\!\widetilde{B}}_6 +
2 m_N^2 x^2 \;\widetilde{\!\widetilde{B}}_8)(x) \nnb \\
%
%
\ek {4\over x} e_{q_3} \Bigg\{ m_N^2 (1+\beta)\Big[ m_N x (m_N^2 x^2 + Q^2)   
\; \widehat{\!\widehat{B}}_6 - m_{q_3} (Q^2 \; \widehat{\!\widehat{B}}_6 + 2 x^2
m_N^2 \; \widehat{\!\widehat{B}}_8) \Big] \nnb \\
\ar \Big[x^2 m_{q_3} m_N^4(1-\beta) (\; \widehat{\!\widehat{C}}_6 - \;\widehat{\!\widehat{D}}_6)
\Big] \Bigg\} (x)~. \nnb
\eea
%
%
%
\section{Structure $q_\mu \rlap/{q}$}
%
%
%
\bea
\rho_{3,1} (x)\es 0~, \nnb \\ \nnb \\
%
%
\rho_{3,2} (x)\es
%
%
{1 \over x} e_{q_1} m_N^2 \Bigg\{ 4 (1+\beta) \, \check{\!\check{B}}_6 (x) \nnb \\
\ek x \Big[ 2 (1+\beta) (\check{B}_2
+ 2 \check{B}_4 + 3 \check{B}_5 + 6 \check{B}_7 - \check{E}_1 + \check{H}_1) \nnb \\
\ar (1-\beta) (2 \check{C}_2 + \check{C}_4 -3 \check{C}_5 + 2
\check{D}_2 - \check{D}_4 + 3 \check{D}_5) \Big]\Bigg\} (x) \nnb \\
%
%
\ek e_{q_2} m_N^2  \Big[2 (1+\beta) (\widetilde{B}_2 +                    
\widetilde{B}_4 + 2 \widetilde{B}_5 + 4 \widetilde{B}_7)
+ (1-\beta) (\widetilde{C}_4 - \widetilde{C}_5 
+ \widetilde{D}_4 - \widetilde{D}_5) \Big] (x) \nnb \\
%
%
%
\ar {2\over x} e_{q_3} m_N^2 \Bigg\{ 2 (1+\beta) \; \widehat{\!\widehat{B}}_6 \nnb \\
\ek x \Big[ (1+\beta) (\widehat{B}_4 + \widehat{B}_5 + 2        
\widehat{B}_7 - \widehat{E}_1 + \widehat{H}_1) +
(1-\beta) (\widehat{C}_2 - \widehat{C}_5 + \widehat{D}_2 + \widehat{D}_5)
\Big] \Bigg\} (x)~, \nnb \\ \nnb \\ 
%
%
\rho_{3,3} (x)\es
%
%
{4\over x} e_{q_1} m_N^2 (1+\beta) (m_N^2 x^2 +Q^2) \, \check{\!\check{B}}_6 (x)\nnb \\
%
%
\ar 4 e_{q_2} m_N^3 \Big[ (1+\beta)           
(\; \widetilde{\!\widetilde{C}}_6 + \; \widetilde{\!\widetilde{D}}_6) 
+ (1-\beta) (\; \widetilde{\!\widetilde{B}}_6 - 2 \;
\widetilde{\!\widetilde{B}}_8) \Big] (x) \nnb \\
%
%
\ar {4\over x} e_{q_3} m_N^2 \Bigg\{ (1+\beta)\Big[ (m_N^2 x^2 +
Q^2) \; \widehat{\!\widehat{B}}_6 + x m_{q_3} m_N (\; \widehat{\!\widehat{B}}_6 -
2 \; \widehat{\!\widehat{B}}_8) \Big] \nnb \\
\ar x m_{q_3} m_N (1-\beta)
(\;\widehat{\!\widehat{C}}_6 - \;\widehat{\!\widehat{D}}_6) \Big] \Bigg\}
(x)~. \nnb
\eea
%
%
%
%
Explicit forms of the functions  ${\cal F}(x_i)$ appearing in the
expressions for $\rho_{i,n}$ are defined in the following way:

\bea
\label{eslff12}
\check{\cal F}(x_1) \es \int_1^{x_1}\!\!dx_1^{'}\int_0^{1- x^{'}_{1}}\!\!dx_3\,
{\cal F}(x_1^{'},1-x_1^{'}-x_3,x_3)~, \nnb \\
\check{\!\!\!\;\check{\cal F}}(x_1) \es 
\int_1^{x_1}\!\!dx_1^{'}\int_1^{x^{'}_{1}}\!\!dx_1^{''}
\int_0^{1- x^{''}_{1}}\!\!dx_3\,
{\cal F}(x_1^{''},1-x_1^{''}-x_3,x_3)~, \nnb \\
\widetilde{\cal F}(x_2) \es \int_1^{x_2}\!\!dx_2^{'}\int_0^{1- x^{'}_{2}}\!\!dx_1\,
{\cal F}(x_1,x_2^{'},1-x_1-x_2^{'})~, \nnb \\
\widetilde{\!\widetilde{\cal F}}(x_2) \es 
\int_1^{x_2}\!\!dx_2^{'}\int_1^{x^{'}_{2}}\!\!dx_2^{''}
\int_0^{1- x^{''}_{2}}\!\!dx_1\,
{\cal F}(x_1,x_2^{''},1-x_1-x_2^{''})~, \nnb \\
\widehat{\cal F}(x_3) \es \int_1^{x_3}\!\!dx_3^{'}\int_0^{1- x^{'}_{3}}\!\!dx_1\,
{\cal F}(x_1,1-x_1-x_3^{'},x_3^{'})~, \nnb \\
\widehat{\!\widehat{\cal F}}(x_3) \es 
\int_1^{x_3}\!\!dx_3^{'}\int_1^{x^{'}_{3}}\!\!dx_3^{''}
\int_0^{1- x^{''}_{3}}\!\!dx_1\,
{\cal F}(x_1,1-x_1-x_3^{''},x_3^{''})~.\nnb
\eea

Definitions of the functions $B_i$, $C_i$, $D)i$, $E_1$ and $H_1$
which represent the linear combinations of the DAs are given as follows:

\bea
\label{eslff13}
B_2 \es T_1+T_2-2 T_3~, \nnb \\
B_4 \es T_1-T_2-2 T_7~, \nnb \\
B_5 \es - T_1+T_5+2 T_8~, \nnb \\
B_6 \es 2 T_1-2 T_3-2 T_4+2 T_5+2 T_7+2 T_8~, \nnb \\
B_7 \es T_7-T_8~, \nnb \\
B_8 \es  -T_1+T_2+T_5-T_6+2 T_7+2T_8~, \nnb \\
C_2 \es V_1-V_2-V_3~, \nnb \\
C_4 \es -2V_1+V_3+V_4+2V_5~, \nnb \\
C_5 \es V_4-V_3~, \nnb \\
C_6 \es -V_1+V_2+V_3+V_4+V_5-V_6~, \nnb \\
D_2 \es -A_1+A_2-A_3~, \nnb \\
D_4 \es -2A_1-A_3-A_4+2A_5~, \nnb \\
D_5 \es A_3-A_4~, \nnb \\
D_6 \es A_1-A_2+A_3+A_4-A_5+A_6~, \nnb \\
E_1 \es S_1-S_2~, \nnb \\
H_1 \es P_2-P_1~. \nnb
\eea

\newpage

\section*{Figure captions}
{\bf Fig. 1} The dependence of the helicity amplitude $A_{1/2}$
on $Q^2$ at $M^2=2.0~\mbox{GeV}^2$, at several fixed values of the
auxiliary parameter $\beta$. \\ \\
{\bf Fig. 2} The same as in Fig. 1, but for the helicity amplitude
$S_{1/2}$.

\newpage

\begin{figure}
\vskip 3. cm
    \includegraphics{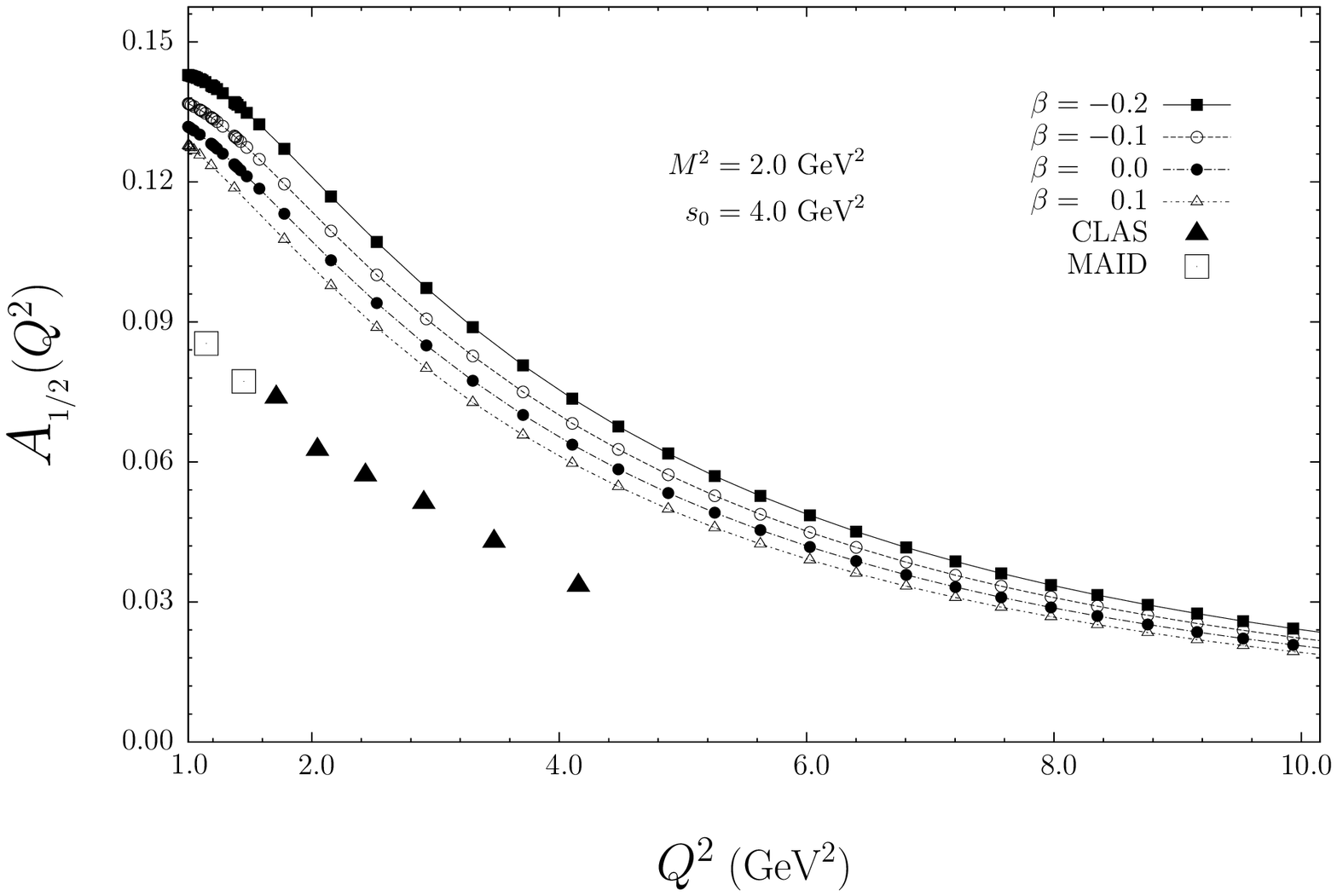}
\vskip 7.0cm
\caption{}
\end{figure}

\begin{figure}
\vskip 3. cm
    \includegraphics{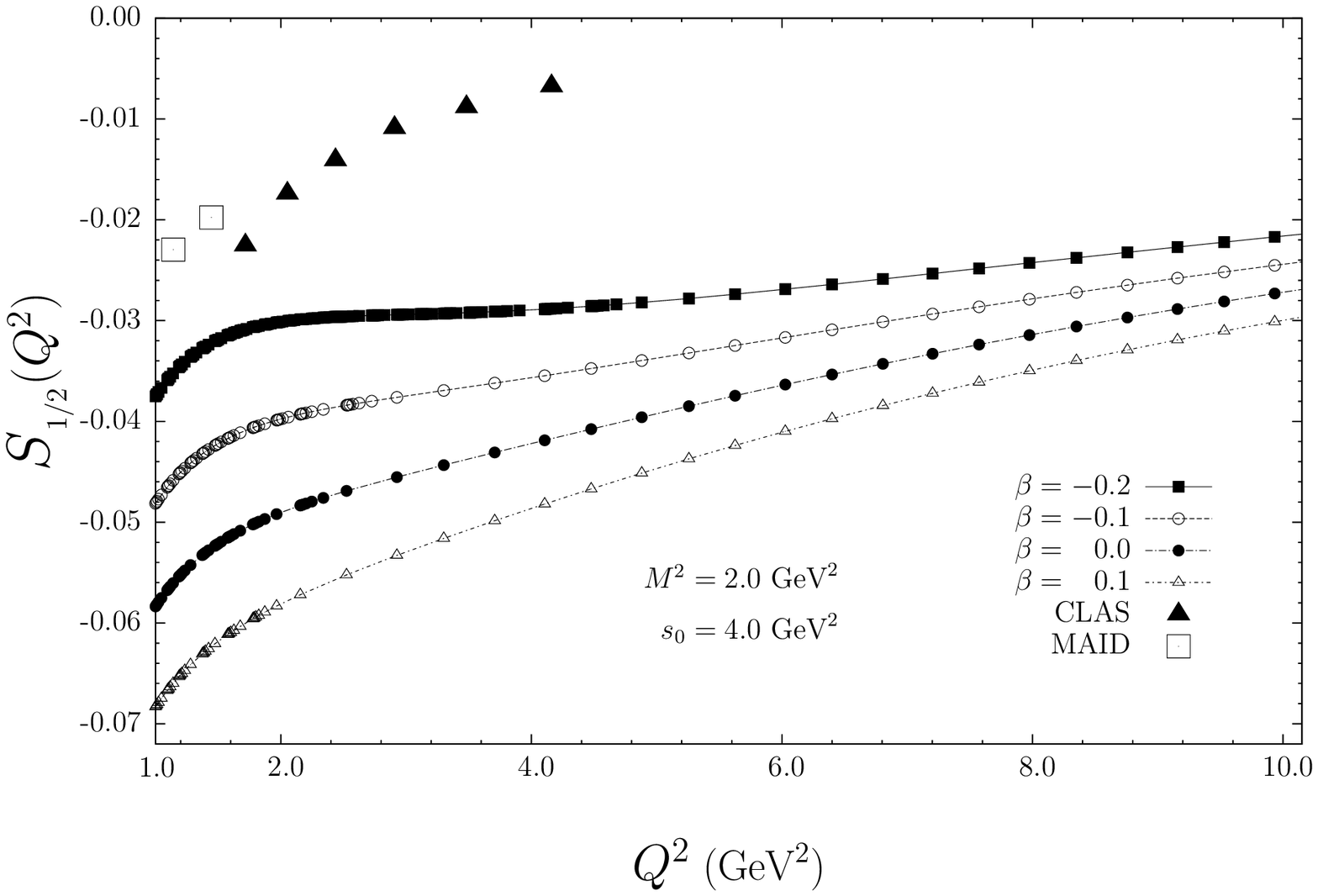}
\vskip 7.0cm
\caption{}
\end{figure}


\newpage

\begin{thebibliography}{99}

\bibitem{Rgsn01} V. M. Braun, A. Lenz, N. Mahnke, and E. Stein,
  Phys. Rev. D {\bf 65}, 074011 (2002).

\bibitem{Rgsn02} I. G. Aznauryan {\it et. al},
  Int. J. Mod. Phys. E {\bf 22}, 1330015 (2013).

\bibitem{Rgsn03} G. Ramalho,
  Phys. Rev. D {\bf 84}, 033007 (2011).

\bibitem{Rgsn04} Huey-Wen Lin, S. D. Cohen, R. G. Edwards, and D. G. Richards,
  Phys. Rev. D {\bf 78}, 114508 (2008).

\bibitem{Rgsn05} V. M. Braun, R. J. Fries, N. Mahnke, and E. Stein,
  Nucl.Phys. B {\bf 589}, 381 (2000);
  Nucl.Phys. B {\bf 607}, 433(E) (2001).

\bibitem{Rgsn06} T. M. Aliev, A. \"{O}zpineci, M. Savc{\i},
  Phys. Rev. D {\bf 66}, 016002 (2002).

\bibitem{Rgsn07} I. G. Aznauryan, V. D. Burkert, and T. S. Lee,
  arXiv: 0810.0997.

\bibitem{Rgsn08} V. M. Braun {\it et. al},
  Phys. Rev. Lett. {\bf 103}, 072001 (2009).

\bibitem{Rgsn09} G. Duplancic, A. Khodjamirian, Th. Mannel, B. Melic, and N. Offen, 
  JHEP {\bf 0804}, 014 (2008).

\bibitem{Rgsn10} I. G. Aznauryan {\it et. al}, (CLAS Collaboration),
  Phys. Rev. C {\bf 80} 055203 (2009).

\bibitem{Rgsn11} H. Denizli, {\it et. al}, CLAS Collaboration,
  Phys. Rev. C {\bf 76}, 015204 (2007).

\bibitem{Rgsn12} D. Drechsel, S. S. Kamalov, and L. Tiator,
  Eur. Phys. J. A {\bf 34}, 69 (2007).

\end{thebibliography}
\end{document}